\documentclass[twocolumn]{aastex631}

\newcommand{\teff}{\mbox{$T_{\mathrm{eff}}$}} 
\newcommand{\logg}{\mbox{log $g$}}

\newcommand{\kzz}{\mbox{$K_{\mathrm{zz}}$}} 

\newcommand{\mic}{\mbox{$\mu$m}}
\newcommand{\wnameM}{WISE~2102$-$44}
\newcommand{\wfnameM}{WISE~J210200.15$-$442919.5}
\newcommand{\wnameP}{WISE~2056+14}
\newcommand{\wfnameP}{WISEPC~J205628.90+145953.3}
\newcommand{\wnameOld}{WISE~0359$-$54}

\begin{document}

\title{A Tale of Two Molecules: The Underprediction of CO$_2$ and Overprediction of PH$_3$ in Late T and Y Dwarf Atmospheric Models}
\shorttitle{A Tale of Two Molecules}
\author[0000-0002-6721-1844]{Samuel A. Beiler}
\affiliation{Ritter Astrophysical Research Center, Department of Physics \& Astronomy,
University of Toledo, 2801 W. Bancroft St.,
Toledo, OH 43606, USA}

\author[0000-0003-1622-1302]{Sagnick Mukherjee}
\affiliation{Department of Astronomy and Astrophysics, University of California, Santa Cruz, 1156 High Street, Santa Cruz, CA 95064, USA}
\affiliation{Department of Physics and Astronomy, Johns Hopkins University, Baltimore, MD, USA \\ }

\author[0000-0001-7780-3352]{Michael C. Cushing}
\affiliation{Ritter Astrophysical Research Center, Department of Physics \& Astronomy,
University of Toledo, 2801 W. Bancroft St.,
Toledo, OH 43606, USA}

\author[0000-0002-6721-1844]{J. Davy Kirkpatrick}
\affiliation{IPAC, Mail Code 100-22, Caltech, 1200 E. California Boulevard, Pasadena, CA 91125, USA}

\author[0000-0003-4269-260X]{Adam C. Schneider}
\affiliation{United States Naval Observatory, Flagstaff Station, 10391 West Naval Observatory Road, Flagstaff, AZ 86005, USA}

\author[0009-0009-4489-0192]{Harshil Kothari}
\affiliation{Ritter Astrophysical Research Center, Department of Physics \& Astronomy,
University of Toledo, 2801 W. Bancroft St.,
Toledo, OH 43606, USA}

\author[0000-0002-5251-2943]{Mark S. Marley}
\affiliation{Lunar and Planetary Laboratory, University of Arizona, 1629 E. University Boulevard, Tucson, AZ 85721, USA}

\author[0000-0001-6627-6067]{Channon Visscher}
\affiliation{Department of Chemistry and Planetary Sciences, Dordt University, Sioux Center, IA, USA}
\affiliation{Center for Exoplanetary Systems, Space Science Institute, Boulder, CO, USA}

\begin{abstract}
The sensitivity and spectral coverage of JWST is enabling us to test our assumptions of ultracool dwarf atmospheric chemistry, especially with regards to the abundances of phosphine (PH$_3$) and carbon dioxide (CO$_2$). In this paper, we use NIRSpec PRISM spectra ($\sim$0.8$-$5.5 \mic, $R\sim$100) of four late T and Y dwarfs to show that standard substellar atmosphere models have difficulty replicating the 4.1$-$4.4 \mic{} wavelength range as they predict an overabundance of phosphine and an underabundance of carbon dioxide. To help quantify this discrepancy, we generate a grid of models using PICASO based on the Elf Owl chemical and temperature profiles where we include the abundances of these two molecules as parameters. The fits to these PICASO models show a consistent preference for orders of magnitude higher CO$_2$ abundances and a reduction in PH$_3$ abundance as compared to the nominal models. This tendency means that the claimed phosphine detection in UNCOVER$-$BD$-$3 could instead be explained by a CO$_2$ abundance in excess of standard atmospheric model predictions; however the signal-to-noise of the spectrum is not high enough to discriminate between these cases. We discuss atmospheric mechanisms that could explain the observed underabundance of PH$_3$ and overabundance of CO$_2$, including a vertical eddy diffusion coefficient (\kzz) that varies with altitude, incorrect chemical pathways, or elements condensing out in forms such as NH$_4$H$_2$PO$_4$. However, our favored explanation for the required CO$_2$ enhancement is that the quench approximation does not accurately predict the CO$_2$ abundance, as CO$_2$ remains in chemical equilibrium with CO after CO quenches.
\end{abstract}


\section{Introduction} \label{sec:intro}
As we enter the 3rd year since the James Webb Space Telescope's (JWST) launch \citep{rigby_science_2023}, we are starting to gain a new perspective on the atmospheres of ultracool objects. In particular, we have seen some of the first JWST observations of the 3.5$-$6 \mic{} region, where the spectral energy distributions of late T and Y dwarfs peak \citep[e.g.][]{burgasser_uncover_2023,luhman_jwstnirspec_2023,lew_high-precision_2024}. This region, which we will refer to as the 5 \mic{} peak, is a general feature of cool substellar objects \citep{marley_atmospheric_1996}. This region contains strong absorption bands of several molecules, including $\mathrm{H}_2\mathrm{O}$, $\mathrm{CH}_4$, $\mathrm{CO}$, and $\mathrm{CO}_2$ and has a long pathlength through the atmosphere owing to low background opacity. This long pathlength allows lower abundance molecules to be observed in this portion of the spectrum, for example, the detection of CO, PH$_3$, and GeH$_4$ in Jupiter \citep{beer_detection_1975,ridgway_800_1976,fink_germane_1978}. As a result, the spectral morphology of this region is especially sensitive to variations in atmospheric abundances and chemistry. 

Atmospheric mixing is a key factor in the chemistry of these cold atmospheres \citep{barshay_chemical_1978,saumon_molecular_2000}. Mixing can keep the atmosphere from achieving chemical equilibrium if the mixing timescale is shorter than the chemical timescale. In late T and Y dwarfs, for example, CH$_4$ is chemically favored in the upper, cooler layers of atmospheres over CO, but CO is still observed because it is being dredged up from deeper, hotter layers at a faster rate than it can be converted to CH$_4$ \citep{lodders_atmospheric_2002,visscher_quenching_2011}. Disequilibrium chemistry has often been observed in brown dwarfs across several molecules \citep[e.g.][]{noll_detection_1997,sorahana_akari_2012,miles_observations_2020}, and is an important consideration in the modeling of their atmospheres. As such, most atmospheric models include disequilibrium chemistry via vertical mixing, parameterized by a vertical eddy diffusion coefficient \kzz{} \citep[e.g.][]{hubeny_systematic_2007,saumon_evolution_2008,phillips_new_2020,mukherjee_probing_2022,lacy_self-consistent_2023}, with a higher value corresponding to stronger mixing. 

Models that include disequilibrium chemistry generally fit observations well, but they still have difficulty fitting the 5 \mic{} peak \citep{beiler_first_2023,luhman_jwstnirspec_2023,leggett_first_2023}. This is in part due to the sheer number of molecular bands present whose abundances impact this portion of the spectrum. In this paper, we show that the two molecules that are causing the most problems for the forward model fits are CO$_2$ and PH$_3$. The models generally predict significant phosphine absorption, and minimal carbon dioxide absorption, which is the opposite of what has been seen in observations \citep{beiler_first_2023,luhman_jwstnirspec_2023}. The absence of observed PH$_3$ absorption in late-type brown dwarfs (with the exception of \citet{burgasser_uncover_2023} claiming a phosphine detection in the galactic disk object UNCOVER$-$BD$-$3) is especially strange given its strong features in the spectra Jupiter and Saturn \citep{ridgway_800_1976,prinn_phosphine_1975,gillett_75-_1974}. 

Previous work has noted that the CO$_2$ absorption band seen in AKARI spectra of late-type T dwarfs is poorly fit by atmospheric models \citep{yamamura_akari_2010}. The CO$_2$ absorption band at 4.2 \mic{} implies an abundance that disequilibrium models could not replicate while simultaneously fitting the CO and CH$_4$ features. \citet{tsuji_akari_2011} and \citet{sorahana_akari_2014} both proposed that enhanced C and O abundances could account for this mismatch. However, after the completion of the AKARI mission in 2011, the CO$_2$ features were unobservable until JWST began observation in 2022. As we enter the era of JWST, understanding the chemistry of CO$_2$ and PH$_3$ is of increased importance.

In this paper, we demonstrate how a variety of standard model grids over-predict and under-predict PH$_3$ and CO$_2$ abundances, respectively, and explore how modifying the abundances of PH$_3$ and CO$_2$ affect the model fits. In $\S$\ref{sec:obvs} we discuss the sample selection and observations, whose reduction we discuss in $\S$\ref{sec:reduc}. In $\S$\ref{sec:fits} we discuss fitting three commonly used model grids to the sample and the difficulties in fitting the 5 \mic{} peak. In $\S$\ref{sec:picaso} we detail the generation of a grid of models which include parameters to modify the abundances of phosphine and carbon dioxide, and compare how this grid of models fit the observations. In $\S$\ref{sec:explain} we discuss what atmospheric processes could result in these modified CO$_2$ and PH$_3$ abundances.

\section{Sample and Observations} \label{sec:obvs}
Our sample includes two previously published JWST Near Infrared Spectrograph \citep[NIRSpec,][]{boker_-orbit_2023} PRISM spectra of WISE~J035934.06$-$540154.6 \citep[Y0,][]{beiler_first_2023} and UNCOVER$-$BD$-$3 \citep[T8 or T9,][]{burgasser_uncover_2023}, as well as two other ultracool dwarfs to more thoroughly probe the spectral sequence: \wfnameM{} (T9) and \wfnameP{} (Y0), hereafter \wnameM{} and \wnameP{}, respectively, both observed in GO Program \#2302. The properties for these two objects are in Table \ref{tbl:prop}. Even though the WISE objects also have MIRI LRS data available, we include only the NIRSpec PRISM spectra so that a clearer comparison can be made between the WISE objects and UNCOVER$-$BD$-$3, which only has NIRSpec PRISM data.

We observed \wnameP{} and \wnameM{} using JWST's NIRSpec, collecting a low-resolution spectrum for both objects. The instrument was used in fixed-slit mode with the CLEAR/PRISM filter and the S200A1 slit ($0\farcs2\times3\farcs2$). This filter allowed us to obtain a spectrum from 0.6--5.3 \mic{} with a spectral resolving power ($R = {\lambda}/{\Delta \lambda}$) of $\sim$100. The observations were completed with the NRSRAPID readout pattern and a 5 point dither, with both object having a total exposure time of 38.95 s (5 integrations of 7.79 s).


\section{Data Reduction} \label{sec:reduc}
We use the JWST pipeline (Version 1.12.1) for the data reduction, using the 11.17.0 CRDS (Calibration Reference Data System) version and 1132.pmap CRDS context to assign the reference files. We modified the ``extract\_1D'' step of the Stage 3 pipeline to extract from the center of the point spread function as we did in \citet{beiler_first_2023}. We also ignore data shortward of the $Y$ band after the first pixel with negative flux. This occurs at approximately 0.7 \mic{} for \wnameM{} and 0.8 \mic{} for \wnameP.

The pre-flight goal for the precision of the absolute flux calibration of JWST spectra was $\sim10$\% \citep{gordon_james_2022}. We improved the absolute calibration by using Spitzer/IRAC Channel 2 ([4.5], 4.5 \mic) photometry \citep{kirkpatrick_further_2012}, which can be found in Table \ref{tbl:prop}. From these flux densities we calculate the scaling factor needed to convert our spectra and their errors to absolute units of Janskys \citep{reach_absolute_2005, cushing_spitzer_2006}.

\begin{deluxetable}{lccr}
\tablecaption{Astrometric and Photometric Properties \label{tbl:prop}}
\tablehead{
\colhead{Property} &
\colhead{\wnameM} &
\colhead{\wnameP} &
\colhead{Ref.}
}
\startdata
Spectral Type & T9 & Y0 & 1,2\\
Parallax  $ \varpi_\mathrm{abs} $ (mas)& $92.9\pm1.9$&$140.8\pm2.0$&3,4\\
Distance (pc)  & $10.76\pm0.22$&$7.1\pm0.1$& 3,4\\
IRAC [3.6] (mag) & $16.325\pm0.036$& $16.031\pm0.030$& 3,3\\
IRAC [4.5] (mag) & $14.223\pm0.019$& $13.923\pm0.018$& 3,3\\
\enddata
\tablerefs{
(1)~\citet{kirkpatrick_further_2012},(2)~\citet{cushing_discovery_2011}, 
(3)~\citet{kirkpatrick_preliminary_2019}, (4)~\citet{tinney_luminosities_2014}.}
\end{deluxetable}

\section{Comparing the Spectra to Standard Atmospheric Models}\label{sec:fits}
We fit this sample of NIRSpec spectra with a variety of standard atmospheric models to demonstrate the difficulty in replicating the spectra at the 5 \mic{} peak. We used the Elf Owl \citep{mukherjee_sonora_2024}, ATMO2020++ \citep{phillips_new_2020,meisner_exploring_2023}, and LOW-Z models \citep{meisner_new_2021}, all of which include disequilibrium chemistry. The Elf Owl and LOW-Z models parameterize this with a constant vertical eddy diffusion coefficient (\kzz). The ATMO2020++ models do not have an independent \kzz{} parameter, but instead defines a vertically uniform \kzz{} as a function of \logg{} [$\mathrm{cm~s}^{-2}$] such that log \kzz$= 5+2\times(5-$\logg) [$\mathrm{cm}^2\ \mathrm{s}^{-1}$]. The parameter ranges we use for each model grid can be found in Table \ref{tbl:models}. The increments on the \logg{} and \teff{} parameters change both between and within models. The increments for the Elf Owl models can be found in \citet{mukherjee_sonora_2024} (Table 1). The ATMO2020++ model grid increments \teff{} at 25 K from 250-300 K, 50 K from 300-500 K, and 100 K elsewhere, while the LOWZ model grid increments at 50 K below 1000 K, and 100 K above 1000 K. For \logg{}, the ATMO2020++ and LOWZ model grids both increment at 0.5 dex, with the exception of LOWZ including models where \logg=5.25 [cm s$^{-2}$].

\begin{deluxetable*}{llccccrcc}
\tablecaption{Atmospheric Model Parameter Ranges\label{tbl:models}}
\tablehead{
\colhead{Model Name} &
\colhead{\teff{} (K)} &
\colhead{\logg{} [cm s$^{-2}$]} &
\colhead{log \kzz{} [cm$^2$ s$^{-1}$]} &
\colhead{[M/H]} &
\colhead{C/O}
}
\startdata
Sonora Elf Owl& 2400$-$275& 3.25$-$5.5  & 2, 4, 7, 8, 9 & $-$1.0, $-$0.5, 0.0, 0.5, 0.7, 1.0& 0.23, 0.46, 0.92, 1.15 \tablenotemark{a}  \\
ATMO2020++&1200$-$250 & 2.5$-$5.5& $5+2\times(5-$\logg) & $-$1.0, $-$0.5, 0.0 & 1 \\
LOWZ & 1600$-$500 & 3.5$-$5.25 & 0, 2, 10 & $-$2.5 to 1.0& 0.1, 0.55, 0.85 \\
\enddata
\tablenotetext{a}{These values are often labeled as (C/O)/(C/$\mathrm{O}_\odot$), where C/O$_\odot$ = 0.458 (i.e. 0.5, 1, 2, and 2.5), but we have converted them for ease of comparison.}
\end{deluxetable*}

The published version of the Elf Owl model grid handles the phosphine abundance differently than the model's general disequilibrium chemistry scheme. Instead PH$_3$ is assumed to be in chemical equilibrium, due to the non-detection of phosphine in most brown dwarf atmospheres. Since we are interested in the current disequilibrium chemistry paradigm and its prediction of phosphine abundances, we use an earlier version of the Elf Owl models which uses a phosphine abundance consistent with the general disequilibrium chemistry scheme. This model gird is publicly available at \url{https://doi.org/10.5281/zenodo.11370830}.

To compare the models to the observed spectra, we first need to convolve the model spectra to the resolving power of the NIRSpec PRISM spectra at each wavelength due to the non-uniform resolving power of the spectrum, which we calculated as detailed in \citet{beiler_first_2023}. The best-fit model for each grid is the one that minimizes $\chi^2$, defined as:
\begin{equation}
    \chi^2 = \sum_i \left( \frac{f_{\lambda,i}-C M_{\lambda,i}}{\sigma_{\lambda,i}^2} \right) ^2,
\end{equation}
where $M_{\lambda,i}$, $f_{\lambda,i}$, and $\sigma_{\lambda,i}$ are respectively the model flux density, observed flux density, and the observed flux density uncertainty in Janskys at each data point $i$. $C$ is the scaling factor for the model spectrum that minimizes $\chi^2$, defined by:
\begin{equation}\label{eqn:C}
    C=\frac{\sum\limits_{i}M_{\lambda,i} f_{\lambda,i}/\sigma_{\lambda,i}^2}{\sum\limits_{i}{M_{\lambda,i}^2/\sigma_{\lambda,i}^2}}.
\end{equation}

\begin{figure*}
\includegraphics[width=\textwidth]{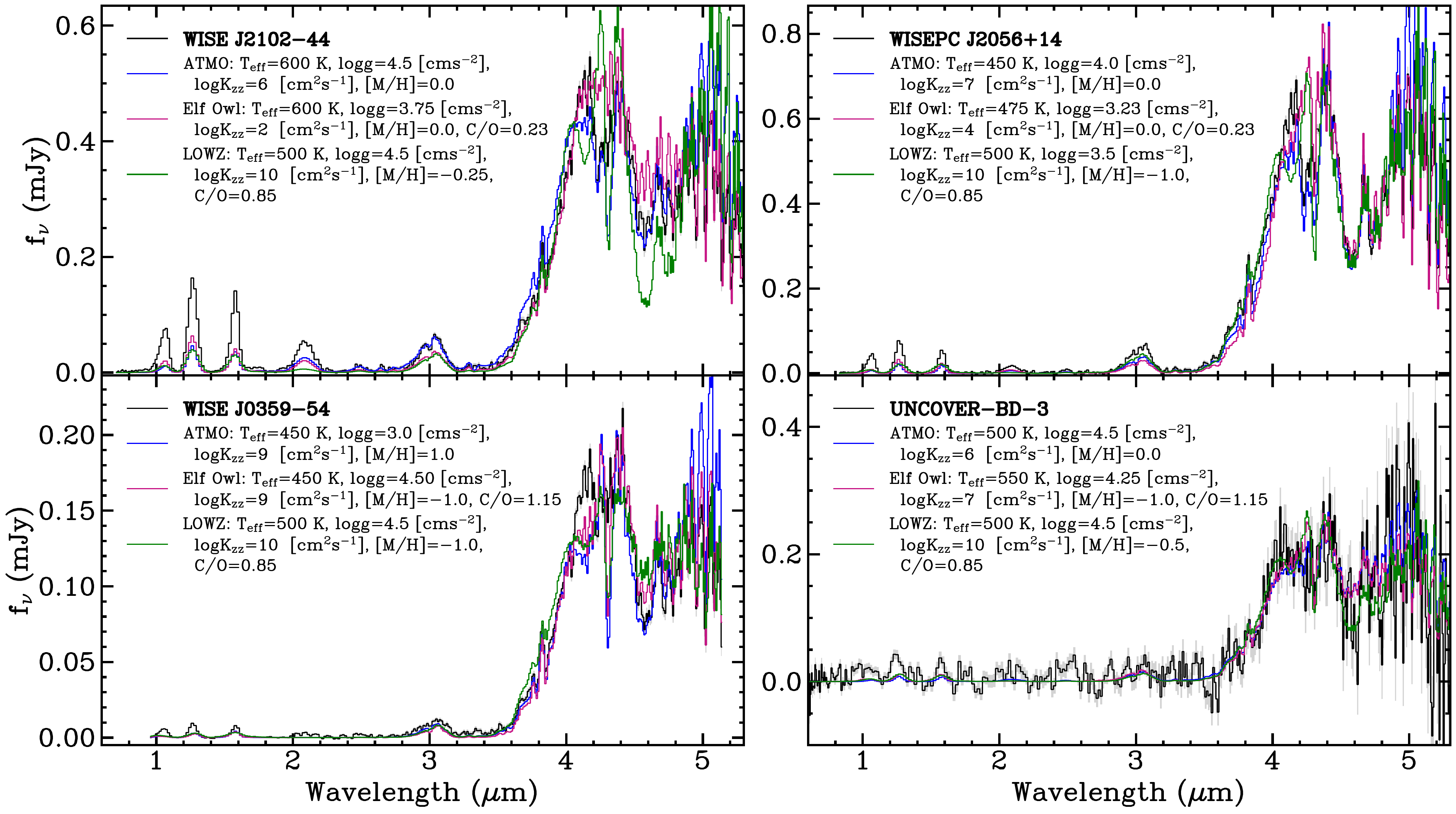}
\centering
\caption{The four objects in this sample (black) plotted alongside the best-fit model for each of the three nominal model grids: ATMO2020++ (blue), Sonora Elf Owl (purple), LOWZ (green). The models generally fit well, but fit the 4--4.5 \mic{} region poorly where PH$_3$ and CO$_2$ affect the spectrum. Uncertainties on the data are shown as grey bars on the data. The two newly published spectra shown in this figure are available as the Data behind the Figure.} \label{fig:modelFits}
\end{figure*}

\begin{figure*}
\includegraphics[width=\textwidth]{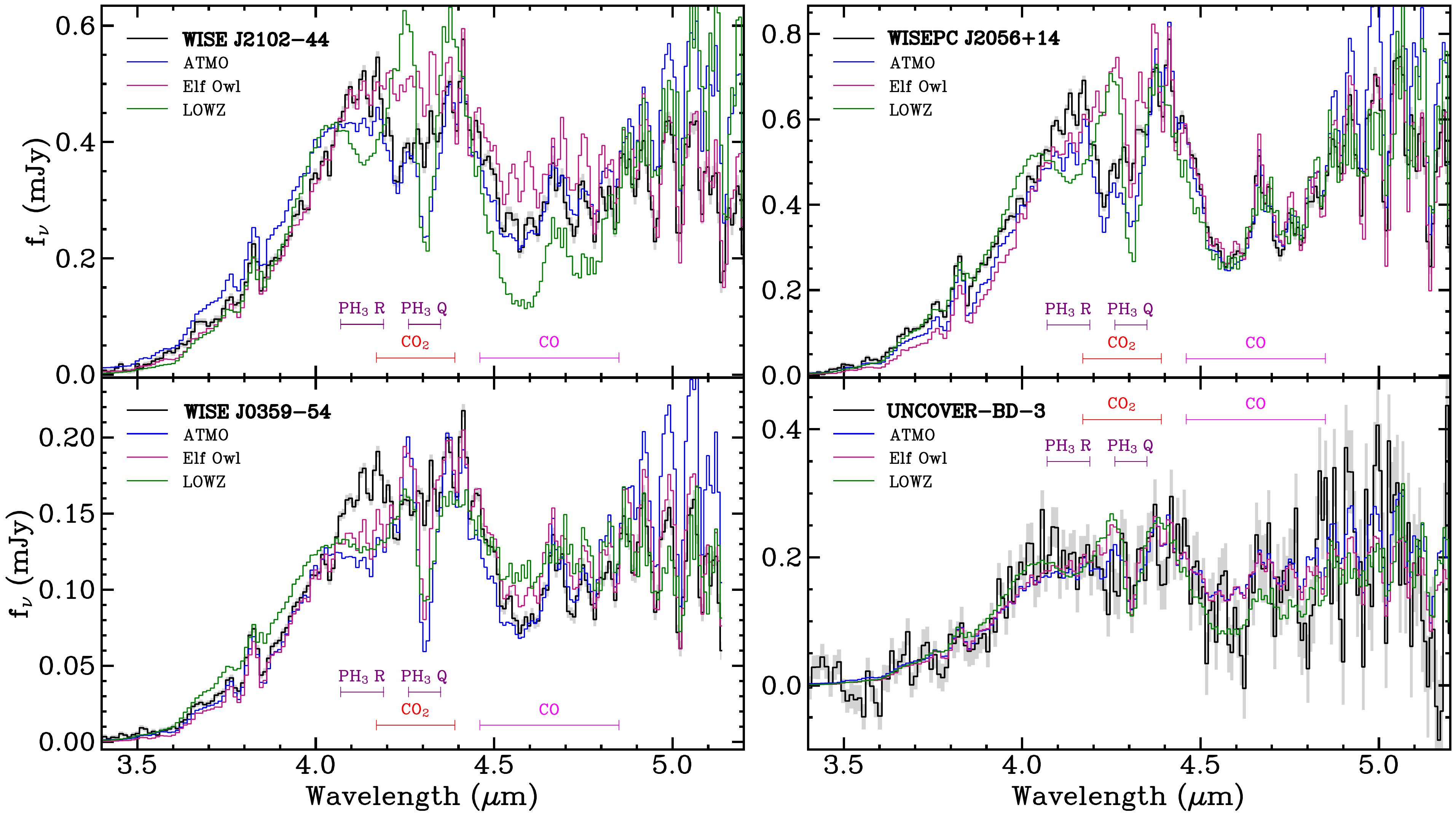}
\centering
\caption{The same plot as Figure \ref{fig:modelFits}, but zoomed in to clearly show the 4--4.5 \mic{} region where PH$_3$ and CO$_2$ affect the spectrum. For all of the models, the PH$_3$ absorption is too strong in at least one of the $Q$ and $R$ branches, and the CO$_2$ is generally too weak.} \label{fig:modelFitsZOOM}
\end{figure*}

The best-fit model parameters for these three nominal model grids are reported in Table \ref{tbl:best}, and Figure \ref{fig:modelFits} shows the best-fit model spectra plotted against all four objects. For UNCOVER--BD--3, we find a different best-fit LOWZ model than \citet{burgasser_uncover_2023} (which is \teff{}=550 K, \logg{}=5.25, [M/H] = $-$0.5), but the $\Delta\chi^2$ between the two best-fit models is only 17 and is likely caused by our choice to convolve the model to NIRSpec's non-uniform resolving power instead of assuming a uniform resolving power. The best-fit models of the three model grids provide reasonable fits to these four objects, with similar temperatures and surface gravities to the other models. The range where the models do have difficulty replicating these observations is from 4--4.5 \mic{} due to predicting strong absorption (see Figure \ref{fig:modelFitsZOOM}, a zoomed in version of Figure \ref{fig:modelFits}).

\begin{deluxetable*}{llcDcDcDr}
\tablecaption{Best-Fit Nominal Atmospheric Model Parameters \label{tbl:best}}
\tablehead{
\colhead{Object Name} &
\colhead{Model Name} &
\colhead{\teff{} (K)} &
\twocolhead{\logg{} [cm s$^{-2}$]} &
\colhead{log \kzz{} [cm$^2$ s$^{-1}$]} &
\twocolhead{[M/H]} &
\colhead{C/O} &
\twocolhead{$\chi^2$} &
\colhead{$\chi^2_r$}
}
\decimals
\startdata
\wnameM&Sonora Elf Owl & 600 & 3.75 & 2   & 0.0   & 0.23 \tablenotemark{a}  & 5206.31 & 13.38 \\
       &ATMO2020++      & 600 & 4.5  & 6 & 0.0   & 0.5  & 10197.29 & 26.21 \\
       &LOWZ           & 500 & 4.5  & 10  & $-$0.25 & 0.85 & 19695.41 & 50.63 \\
\hline
\wnameP&Sonora Elf Owl & 475 & 3.25 & 4   & 0.0    & 0.23 \tablenotemark{a} & 11246.58 & 30.07 \\
       &ATMO2020++      & 450 & 4.0  & 7 & 0.0    & 0.5  & 3881.94 & 10.38\\
       &LOWZ           & 500 & 3.5  & 10  & $-$1.0 & 0.85 & 16547.60 & 44.24 \\
\hline
WISE~0359$-$54&Sonora Elf Owl & 450 & 4.5 & 9   & $-$1.0 & 1.15 \tablenotemark{a} & 4220.04 & 12.34\\
              &ATMO2020++      & 450 & 3.0 & 9 & $-$1.0 & 0.5  & 7661.16 & 22.04\\
              &LOWZ           & 500 & 4.5 & 10  & $-$1.0 & 0.85 & 10455.49 & 30.57\\
\hline
UNCOVER$-$BD$-$3&Sonora Elf Owl & 550 & 4.25 & 7   & $-$1.0 & 1.15 \tablenotemark{a} & 749.66 & 1.86\\
                &ATMO2020++      & 500 & 4.5  & 6 & 0.0    & 0.5  & 764.05 & 1.89\\
                &LOWZ           & 500 & 4.5  & 10  & $-$0.5 & 0.85 & 814.40 & 2.02\\                
\enddata
\tablenotetext{a}{These models report C/O as (C/O)/(C/$\mathrm{O}_\odot$), where C/O$_\odot$ = 0.458, but we have converted them for ease of comparison.}
\end{deluxetable*}

\begin{figure}
\includegraphics[width=0.47\textwidth]{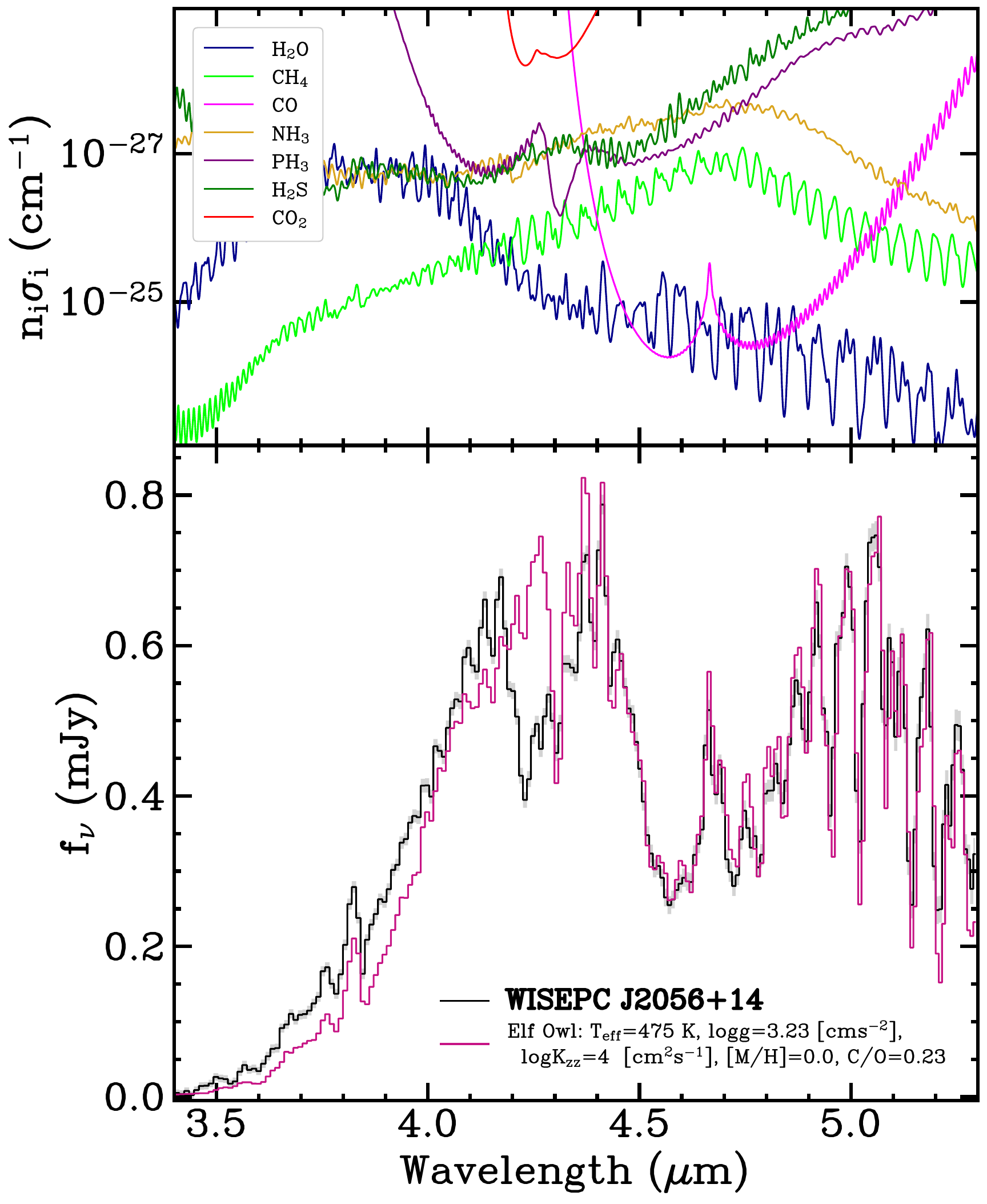}
\centering
\caption{\textbf{Top:} The absorption coefficents ($\alpha_i$=$n_i\sigma_i$) for important molecules in the 3.5--5.3 \mic{} range for important molecules (H$_2$O, CH$_4$, CO, NH$_3$, PH$_3$, H$_2$S, and CO$_2$).  The abundances ($n_i$) are from the best-fit Sonora Elf Owl model for \wnameP{} at P = 1 bar. The cross-sections ($\sigma_i$) are from PICASO at P = 1 bar and T = 725 K, the closest temperature to the best-fit Elf Owl model at 1 bar (T = 745 K). Note the y-axis is inverted for ease of comparison to the spectrum. \textbf{Bottom:} A zoom-in on the 3.4--5.3 \mic{} region of the \wnameP{} portion of Figure \ref{fig:modelFits}, showing clearly how the best-fit Elf Owl model does not fit the CO$_2$ and PH$_3$ features of the observed spectrum.}
\label{fig:2056}
\end{figure}

To get a sense are what is causing the poor fits to this region, in Figure \ref{fig:2056} we plot \wnameP{} alongside the absorption coefficients ($\alpha$=$n\sigma$, where $n$ is the number density and $\sigma_i$ is the cross section) for important molecules (H$_2$O, CH$_4$, CO, NH$_3$, PH$_3$, H$_2$S, and CO$_2$). The number density are from the best-fit Sonora Elf Owl model at $P$ = 1 bar and $T$ = 745 K, and the cross sections are from PICASO at $P$ = 1 bar and the closest temperature available, $T$ = 725 K.

The 5 \mic{} peak is predominately shaped by the $\nu_2$ water band, the $\nu_3$ methane band, and the 1-0 X$^1\Sigma^+$-X$^1\Sigma^+$ band of carbon monoxide. These molecular features are as a whole well fit with some discrepancy in the height of the water features past 4.8 \mic. However, the worst fitting region is 4--4.5 \mic{}, where $\nu_3$ CO$_2$ band and the overlapping $\nu_1$ and $\nu_3$ PH$_3$ bands play a strong role in shaping the spectra. It is clear from these fits that the models have an overabundance of phosphine, seen most clearly in the deep $Q$-branch of phosphine at 4.3 \mic, but also in the $P$- and $R$-branches that bracket the $Q$-branch. 

This overabundance of phosphine arises as an effect of the non-zero \kzz{} values that are required to fit the prominent CO and $\nu_3$ CH$_4$ band at effective temperatures below $\sim$600 K. This enhancement of CO has been seen many times in the spectra of L, T and Y dwarfs and even Jupiter \citep[e.g.][]{beer_detection_1975,noll_detection_1997,oppenheimer_spectrum_1998,sorahana_akari_2012,miles_observations_2020}. For phosphorus, vertical mixing enhances PH$_3$ over oxidized P-bearing species expected to dominate in chemical equilibrium at these temperatures \citep{wang_modeling_2016,visscher_mapping_2020}. PH$_3$ absorption has been seen in the spectra of Jupiter and Saturn \citep{gillett_75-_1974,prinn_phosphine_1975, ridgway_800_1976}, but has not been found outside of our solar system. For example, in the best-fit parameters for \wnameP{} the phosphine abundance at 1 bar increases by $\approx$100 when going from log \kzz{} = 0 to log \kzz{} = 4 [cm$^2$ s$^{-1}$] \citep{mukherjee_sonora_2023}. Unlike CO, PH$_3$ is not expected to appear in the spectra of hotter L and T dwarfs due to the preference for other species such as P$_2$ in the low pressure regions of these objects \citep{visscher_atmospheric_2006}.

Even though the large \kzz{} values that are selected to fit the CO spectral feature result in an overabundance of PH$_3$, these actually improve the $\chi^2$ compared to low \kzz{} values by compensating for the underabundance of CO$_2$ in the models. Like PH$_3$ and CO, the CO$_2$ abundance is set by the \kzz{} parameter, which dredges up CO$_2$ from deeper atmospheric layers. In the case of the Elf Owl best-fit model for \wnameP, the log \kzz{} = 4 [cm$^2$ s$^{-1}$] increases the CO and CO$_2$ abundance over chemical equilibrium by a factor of $\approx 10^{8}$ and $10^{6}$, respectively. However, this increased CO$_2$ abundance is still too low in models to match our sample. 

The lone exception to the standard models over-abundance of phosphine in this sample is the best-fitting Elf Owl model of \wnameM. A combination of a low \kzz{}, higher \teff, and low C/O ratio are required to fit the carbon monoxide and methane features, and this represses the phosphine in the model. However, even in this case we see the underabundance of carbon dioxide in the model. 

\section{Scaling PH$_3$ and CO$_2$ Abundances with PICASO}\label{sec:picaso}
To test the extent to which the phosphine and carbon dioxide abundances are creating the 4--4.5 \mic{} discrepancy, we generate a grid of models where we include the abundances of these molecules as parameters. PH$_3$ and CO$_2$ are both trace species, and so changing their abundances will not have a significant effect on the pressure/temperature $(P/T)$ profile. As such, we generate our new grid by taking $P/T$ and chemical profiles from the Elf Owl models, multiplying the PH$_3$ and CO$_2$ abundance profiles by various factors, and using the PICASO framework \citep[v3.0,][]{batalha_exoplanet_2019,marley_sonora_2021,mukherjee_picaso_2023} to generate emission spectra. We use the same opacities as \citet{mukherjee_sonora_2023} (Table 2) at a native spectral resolution of $R$=60,000 \citep{batalha_resampled_2020}. We will refer to this grid as the PICASO grid hereafter. The parameter range of the grid can be seen in Table \ref{tbl:picaso}. Adding two parameters greatly increases the number of models, so we reduce the run of parameters compared to the Elf Owl range. In total, we generated 86,400 models. These models, convolved to the resolution of our objects, can be found at \url{https://doi.org/10.5281/zenodo.11370830}.

\begin{deluxetable}{lr}
\tablecaption{PICASO Atmospheric Model Parameter Range \label{tbl:picaso}}
\tablehead{
\colhead{Parameter} &
\colhead{Ranges}
}
\startdata
\teff{} (K) & 400--600 (25 K steps)\\
\logg{} [cm s$^{-2}$] & 3.25--5.5 (0.25 steps) \\
log \kzz{} [cm$^2$ s$^{-1}$] & 2, 4, 7, 9 \\
{[}M/H]& $-$1.0, 0.0 \\
C/O& 0.23, 0.46, 1.15 \tablenotemark{a}\\
$\times$ CO$_2$& 1, 50, and 10$^n$ and 5$\times$10$^n$ for n = 2--4 \\
$\times$ PH$_3$& 1, 0.5, 0.2, 0.1, 0.01\\
\enddata
\tablenotetext{a}{These values are often labeled as (C/O)/(C/$\mathrm{O}_\odot$), where C/$\mathrm{O}_\odot$ = 0.458, but we have converted them for ease of comparison.}
\end{deluxetable}

\begin{figure*}
\includegraphics[width=\textwidth]{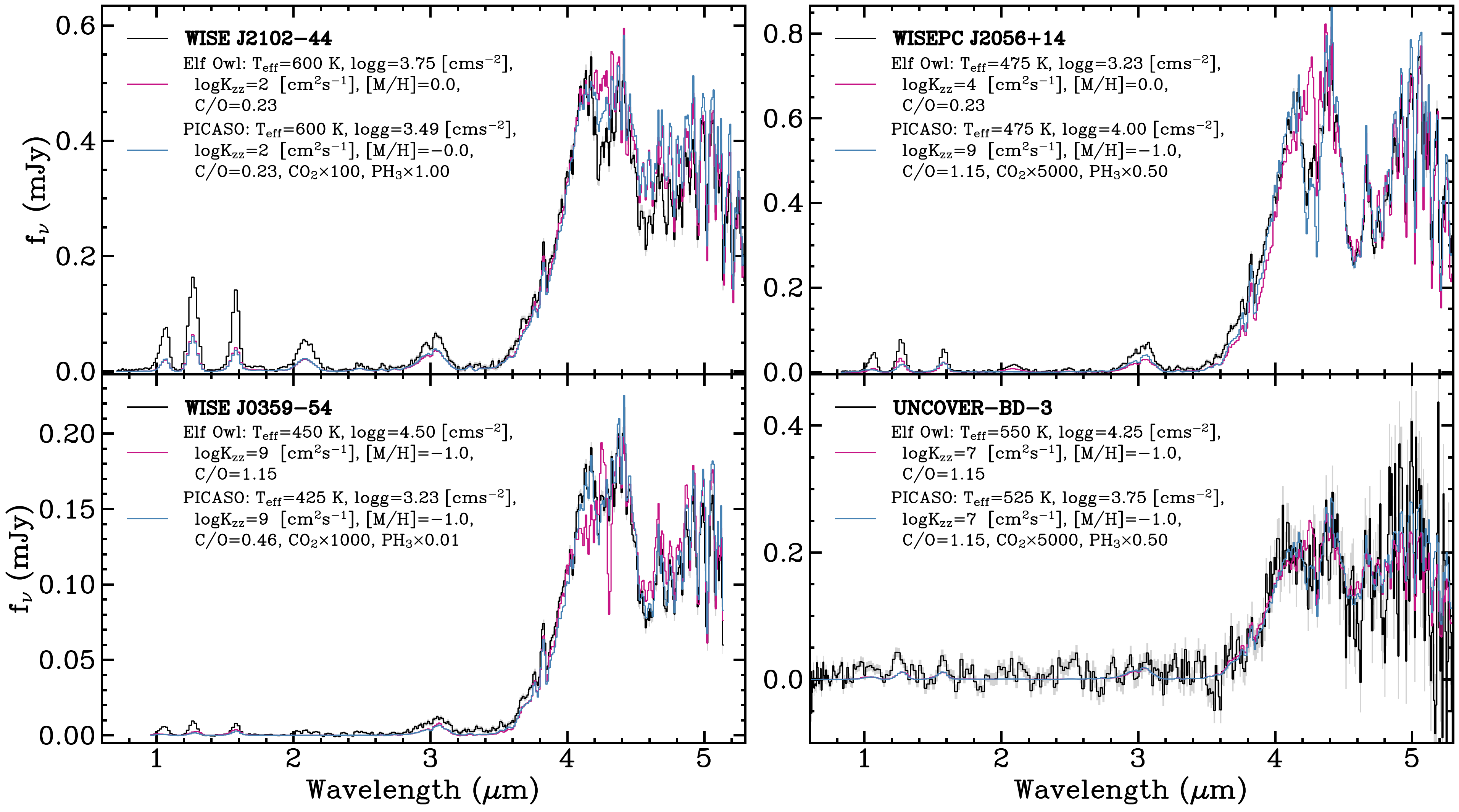}
\centering
\caption{The best-fit Elf Owl (purple) and PICASO (blue) models for the objects in our sample. The PICASO grid, which parameterizes the CO$_2$ and PH$_3$ abundances, grid provides the most consistently good fit across all four objects with a consistent preference for an elevated CO$_2$ abundance, and a decreased PH$_3$ abundance.} \label{fig:modelFitsPicaso}
\end{figure*}

We follow the same fitting procedure as Section \ref{sec:fits}, resulting in the best-fit parameters listed in Table \ref{tbl:bestpicaso} and the best-fit PICASO model spectrum plotted in Figure \ref{fig:modelFitsPicaso} alongside the Elf Owl best-fitting model spectrum. The PICASO models are a significant improvement over the Elf Owl models in every case. The PICASO grid is not the preferred model for every object, as \wnameP{} is better fit by a ATMO2020++ model, but the PICASO grid is a better fit than the other standard models for the rest of the sample. 

\begin{deluxetable*}{lcDcccccDr}
\tablecaption{Best-Fit PICASO Atmospheric Model Parameters \label{tbl:bestpicaso}}
\tablehead{
\colhead{Object Name} &
\colhead{\teff{} (K)} &
\twocolhead{\logg{} [cm s$^{-2}$]} &
\colhead{log \kzz{} [cm$^2$ s$^{-1}$]} &
\colhead{[M/H]} &
\colhead{C/O} &
\colhead{$\times$ CO$_2$} &
\colhead{$\times$ PH$_3$} &
\twocolhead{$\chi^2$} &
\colhead{$\chi^2_r$}
}
\decimals
\startdata
\wnameM          & 600 & 3.5  & 2   & 0.0    & 0.23\tablenotemark{a} & 100  & 1 & 4504.33  & 11.58\\
\wnameP          & 475 & 4    & 9   & $-$1.0 & 1.15\tablenotemark{a} & 5000 & 0.5 & 6457.95 & 17.27\\
WISE~0359$-$54   & 425 & 3.25 & 9   & $-$1.0 & 0.46\tablenotemark{a} & 1000 & 0.01  & 1726.82 & 5.05\\
UNCOVER$-$BD$-$3 & 525 & 3.75 & 7   & $-$1.0 & 1.15\tablenotemark{a} & 5000  & 0.5  & 712.59 & 1.76 \\
\enddata
\tablenotetext{a}{These values are often labeled as (C/O)/(C/O$_\odot$) where C/O$_\odot$ = 0.458, but we have converted them for ease of comparison.}
\end{deluxetable*}

For all of the objects, there is a strong preference for elevated CO$_2$ abundances, up to a factor of 5000 above the base Elf Owl model with the same parameters. This abundance is high enough that the $\nu_3$ CO$_2$ band is now one of the dominant absorption features at 4.25 \mic{} matching observations, which can be seen in Figure \ref{fig:2056Picaso} and \ref{fig:UNCOVERPicaso}. Similar figures for the other two objects can be found in the Appendix. With such a large increase in CO$_2$ abundance, it might be expected that the C/O ratio would be impacted, however for the best-fit models it changes by less than 0.1\% due to CO$_2$ being at such low overall abundances.

\begin{figure}
\includegraphics[width=0.5\textwidth]{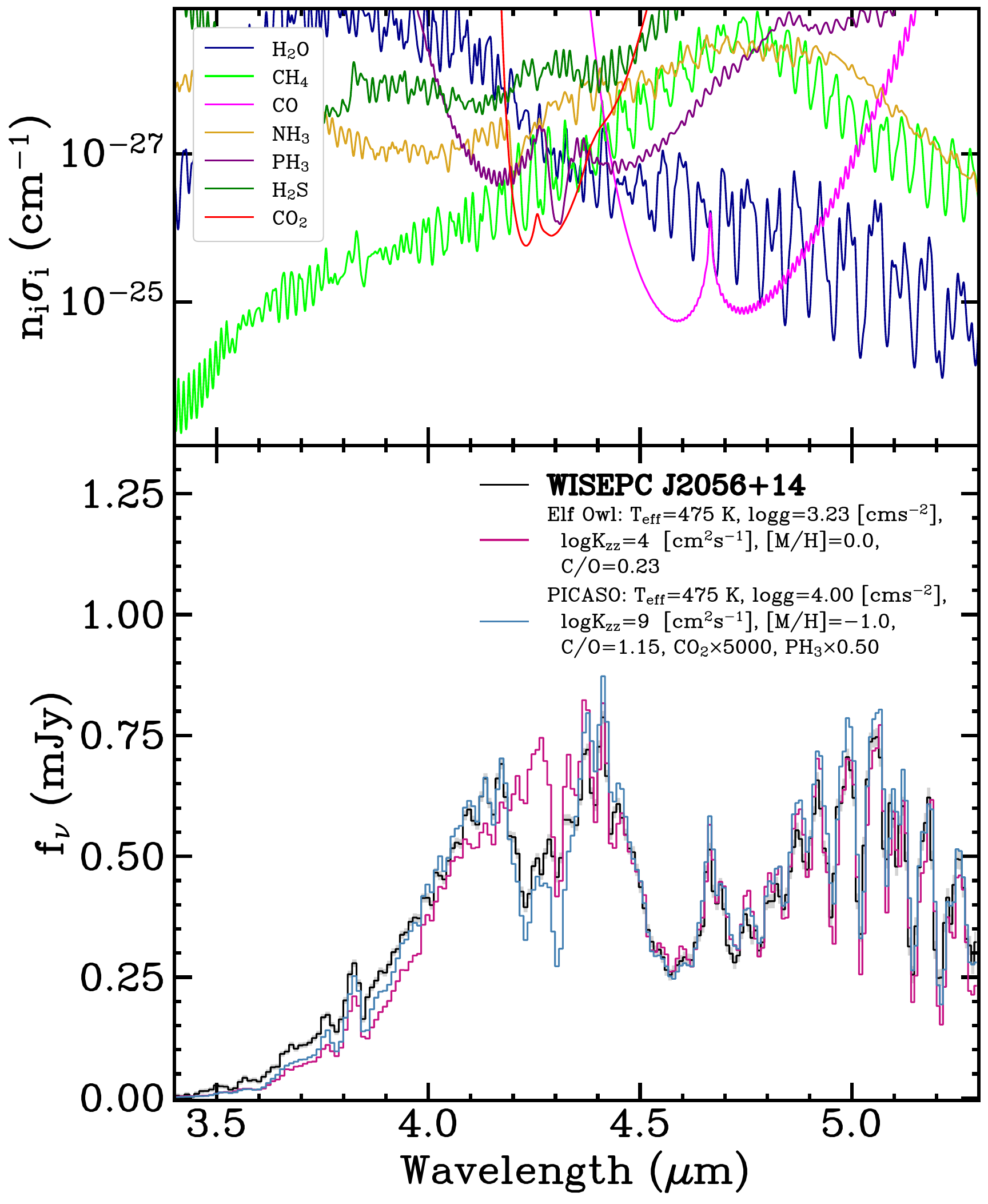}
\centering
\caption{\textbf{Top:} The same plot as Figure \ref{fig:2056}, but this time using the abundances from the PICASO best-fit model at P = 1 bar, where T = 469 K, and the cross-sections at T = 475 K. Note the dramatic increase in the CO$_2$ absorption coefficent. \textbf{Bottom:} The best-fit Elf Owl (purple) and PICASO(blue) models from 3.4--5.3 \mic{}. Note the particular improvement of the fit in the range that CO$_2$ and PH$_3$ dominate (4.1--4.4 \mic).}\label{fig:2056Picaso}
\end{figure}

\begin{figure}
\includegraphics[width=0.5\textwidth]{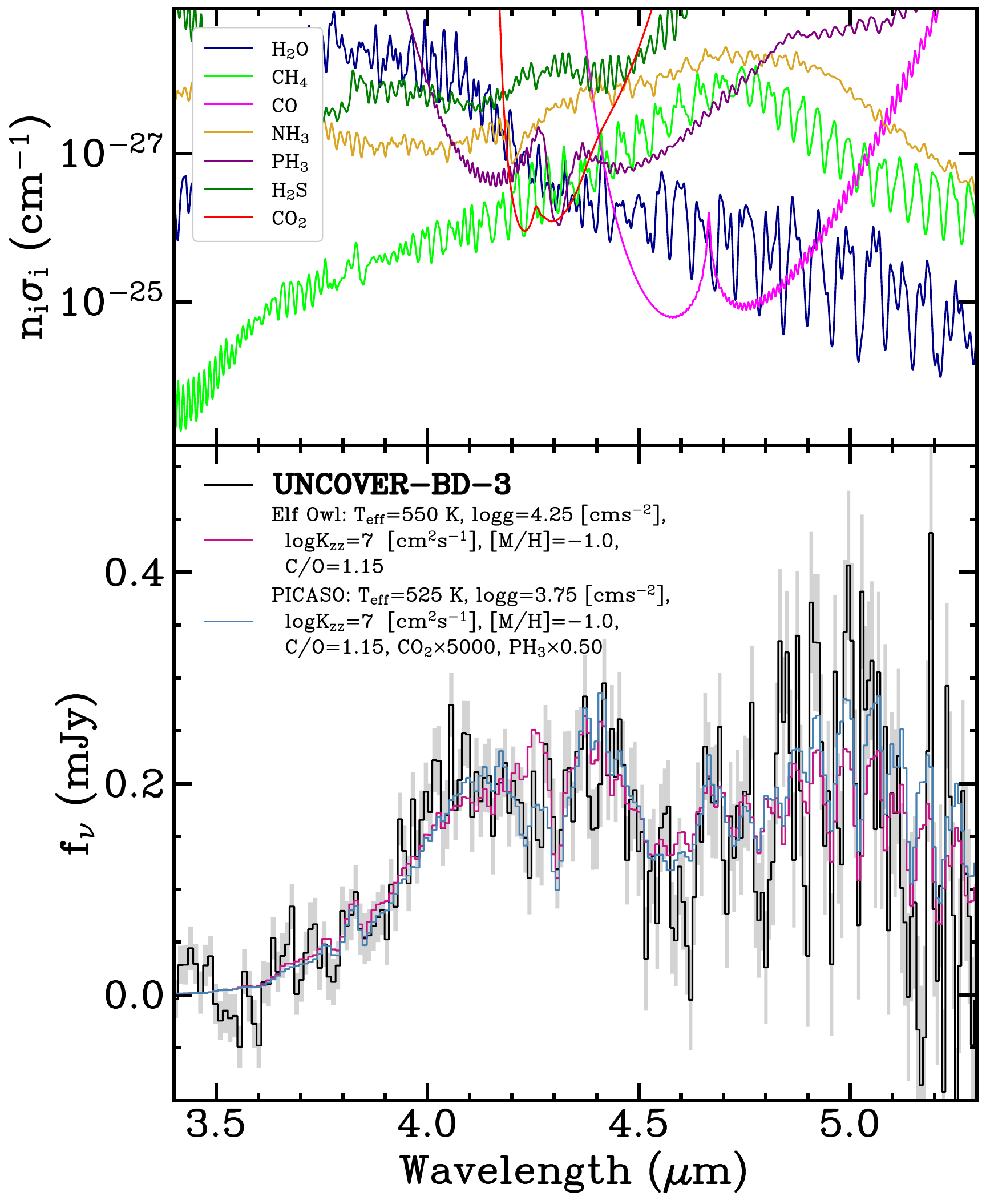}
\centering
\caption{The same plot as Figure \ref{fig:2056Picaso} but with the fits for UNCOVER--BD--3. The abundances are from the PICASO best-fit model at P = 1 bar, where T = 568 K, and the cross sections at T = 575 K.}\label{fig:UNCOVERPicaso}
\end{figure}

We do note that all of the objects which require such high CO$_2$ increases are also best fit by low metallicity PICASO models ([M/H] = $-$1.0). Of these models, only \wnameP{} has solar metallicity models in their top 20 best-fitting models, almost half of which are solar metallicity. To achieve a good fit, these PICASO solar metallicity models all require lower \teff{} (475 or 450 K) and log \kzz{} (4 [cm$^{2}$ s$^{-1}$]) values to try and balance the CO and CH$_4$ abundances, however these models are unfavored as the CH$_4$ absorption is still too strong. All of these factors lead to solar abundance models only requiring a 100-500$\times$ increase of CO$_2$, compared to the 5000$\times$ increase of the sub-solar models. 

All of these objects are best fit with PICASO models whose \logg{}$\leq4$, implying the somewhat unlikely case that they all have low masses (M $<$ 5 $M_\mathrm{Jup}$) and young ages ($\tau<$ 0.3 Gyr) \citet{marley_sonora_2021}. None of the objects have a model with \logg{}$\geq5$ in their 250 best-fit models. This tendency is inherited from the Elf Owl models, whose preference for low gravity fits is seen in this paper, and was also seen in fitting the full spectrum of \wnameOld{} to Elf Owl models \citep{beiler_first_2023}. \citet{kothari_probing_2024} showed that the retrieved gravity and temperature for \wnameOld{} is more in line with the 1-10 Gyr age estimate for field objects, and as such we do not believe the \logg{} values presented here to be accurate. Fortunately, the best-fitting models with \logg{}$>4$ have similar parameters to the overall best-fit models, so while the preference for low gravities is interesting, it does not impact our conclusions.

The PICASO model fits also show a preference for decreasing the abundance of phosphine by a factor of 2-100$\times$ in three of our four objects. The exception is \wnameM, whose phosphine abundance was already too low to be seen in the Elf Owl model for the reasons explained above (high \teff, low \kzz). The change in abundance is not as large for phosphine as it is for carbon dioxide, but it still has a significant impact on the best-fit model spectra. At these lower abundances, the $P$- and $R$-branches of phosphine are no longer discernible in the model spectra. However, while the 4.25-4.4 \mic{} region where the PH$_3$ $Q$-branch sits is now well fit in \wnameOld{} and UNCOVER--BD--3, the best-fit mode1 for \wnameP{} still shows too much PH$_3$ absorption. That said, this region is better reproduced than in the Elf Owl model even with the large increase in \kzz{} that improves the fit redward of 4.6 \mic{}. It might be expected that a further reduction in PH$_3$ would result in a better fit, but without the $P$- and $R$-branches acting to suppress the 4 and 4.4 \mic{} peaks there is too much flux at these peaks to make it the best-fitting model. The 2nd best fit model is identical to the best-fit model, except with $\times$0.2 PH$_3$ instead of $\times$0.5, so the true best fit is likely somewhere between these two values. 

There is a further complication with identifying a PH$_3$ absorption feature at 4.3 \mic. Methane and water both have local peaks that are aligned at this wavelength, which when combined can sometimes be equivalent to or dominate over phosphine in the standard models. This overlap increases the difficulty in identifying phosphine in these objects. With this in mind, it is clear there is no detection of phosphine in the three WISE objects.

It is more difficult to rule out a phosphine detection for UNCOVER--BD--3. \citet{burgasser_uncover_2023} claims the absorption at 4.2 \mic{} is due to the $R$-branch of phosphine. This claim was made assuming the abundances of their LOWZ best-fit model was correct, except for CO$_2$ where they adopted $N_\mathrm{CO_2}/N_\mathrm{CO} = 10^{-3}$ from \citet{yamamura_akari_2010}, which is based off model fits to AKARI spectra (2.6-5 \mic) of mid- and late-T dwarfs. However, we have shown that standard atmospheric models over-predict the phosphine abundance, and the best-fit PICASO model gives $N_\mathrm{CO_2}/N_\mathrm{CO} = 3.8\times10^{-3}$ at 1 bar (T = 567 K). These factors combined to cause CO$_2$ to dominate over PH$_3$ for the best-fit PICASO model. \citet{yamamura_akari_2010} actually estimated $N_{CO_2}/N_{CO} \approx 10^{-3}$ as a lower bound, which would be consistent with our findings and points to enhanced CO$_2$ as another possible explanation for the 4.2 \mic{} absorption feature.

To further demonstrate the lack of clarity in the phosphine abundance in UNCOVER--BD--3, we can look at the 15 best-fitting PICASO models. Of these models there is a roughly even distribution across the phosphine parameter: three models have no change to the phosphine, six have phosphine halved, four have it reduced by a factor of 5, and two have it reduced by a factor of 10. All of these models have roughly the same temperature (525 or 550 K), gravity (3.25$-$4.25 [cm s$^{-2}$]), and CO$_2$ factor (5000$\times$ or 10000$\times$), and all the other parameters are constant. At the signal-to-noise of this spectrum ($\sim8$) it is not possible to definitively determine the abundance of phosphine from models.

With a higher signal-to-noise spectrum of UNCOVER--BD--3 it might be possible to confirm a detection of phosphine; it is the only ultracool object with a published JWST NIRSpec spectrum that potentially has observable phosphine.  However, UNCOVER--BD--3's great distance from Earth makes the exposure time for high signal-to-noise spectrum prohibitive. \citet{burgasser_uncover_2023} predicts that phosphine might be a hallmark of sub-solar metallicity, given that UNCOVER--BD--3 is a disk object, making nearby low-metallicity objects a possibility for confirmation. However, \wnameP{} and \wnameOld{} are best-fit by models with similarly low metallicities, and both lack phosphine absorption features. A JWST Cycle 3 proposal (GO \#4668, PI A. Burgasser) has been accepted to observe 32 L and T subdwarfs, which will help determine if phosphine absorption is a hallmark of sub-solar metallicity.

\section{Possible Explanations for These Differences in Abundances}\label{sec:explain}
There are several possible explanations for the overabundance of phosphine and underabundance of carbon dioxide in standard atmospheric models. For CO$_2$, one possibility is that vertical mixing is function of altitude, as all the models in this paper assume \kzz{} is constant with altitude. The deep atmosphere {\kzz} is expected to set the quench pressures for CH$_4$, CO, and NH$_3$, whereas the upper atmosphere {\kzz} should set the quenched abundance of CO$_2$ \citep{zahnle_methane_2014,mukherjee_probing_2022,mukherjee_sonora_2023,phillips_new_2020}. We perform a test of the effect of {\kzz} varying with altitude where we take a self-consistent Elf Owl model for the best-fit set of parameters and see if a higher or lower {\kzz} in the upper atmosphere can explain the difference in CO$_2$ abundances. We do this test in an ad-hoc manner, where we use the $P/T$ profile of the Elf Owl model, instead of a fully self-consistent approach. We tried to enhance and decrease {\kzz} in the upper atmosphere by a factor relative to the deeper atmosphere {\kzz} and find that changes in the upper atmosphere {\kzz} cannot by themselves explain the much larger enhancements in CO$_2$ abundance required to explain the data.

Another possibility is that our disequilibrium approximations are insufficient for CO$_2$. In the Elf Owl models, the pressure at which a molecule quenches is set by its abundance at the pressure where its chemical timescale becomes slower than the atmospheric mixing timescale. This approximation works well for most molecules, but does not account for the fact that CO quenches at a higher pressure than CO$_2$. After CO quenches with respect to CH$_4$ (by quenching of CO$\rightleftharpoons$CH$_4$ reaction pathways), CO$_2$ remains in equilibrium with the quenched CO abundance due to relatively faster CO$\rightleftharpoons$CO$_2$ reaction pathways. This results in the CO$_2$ abundance \textit{increasing} with altitude above the CO quench level, until CO$_2$ itself quenches to a fixed abundance \citep[][]{visscher_deep_2010}.

\begin{figure*}
\includegraphics[width=0.65\textwidth]{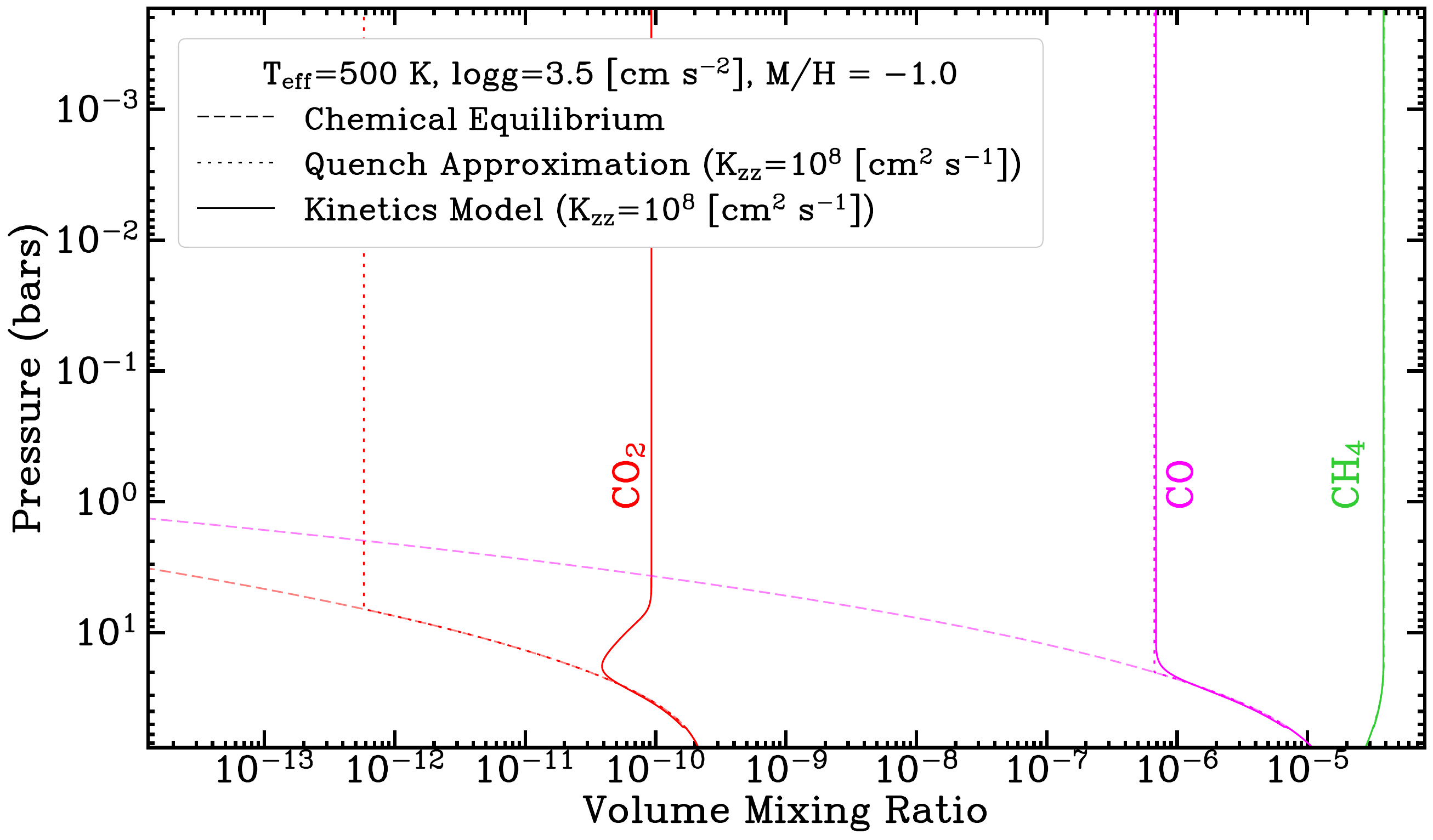}
\centering
\caption{The chemical profiles of CO$_2$ (red), CO (pink), and CH$_4$ (green) assuming various chemical schemes within the Elf Owl framework at 500 K, \logg{} = 4.5 [cm s$^{-1}$], [M/H]=$-1.0$, and solar C/O ratio. The dashed lines are from a model assuming equilibrium chemistry (meaning \kzz=0), the dotted lines are from a model with disequilibrium chemistry implemented via a quenching timescale approximation, and the solid lines are from a kinetics model. Both nonequilibrium models have a log \kzz{} of 8 [cm$^2$ s$^{-1}$]. The kinetic model enhances CO$_2$ by a factor of $\sim$200 over the base Elf Owl disequilibrium models, due to CO$\rightleftharpoons$CO$_2$ reactions keeping CO$_2$ in equilibrium with the quenched CO. This enhancement factor is not as large as the best-fit PICASO enhancement of CO$_2$ ($\sim$1000$\times$), but does move the CO$_2$ meaningfully closer to the preferred value. Due to CO$_2$ being a trace species, CO and CH$_4$ are not strongly impacted by this change.}\label{fig:CO2Profiles}
\end{figure*}

To demonstrate this behavior, we model the CO, CO$_2$, and CH$_4$ abundances with the Elf Owl framework assuming chemical equilibrium, a quench approximation, and a full 1D chemical kinetics model \citep[\textit{Photochem},][]{wogan_jwst_2024} with parameters of 500 K, \logg{}=3.5 [cm s$^{-1}$], [M/H]=$-1.0$, and solar C/O, which roughly represents the best-fitting PICASO models for this sample. These abundance profiles can be seen in Figure \ref{fig:CO2Profiles}. The kinetics and quench approximation models have log \kzz = 8 [cm$^2$ s$^{-1}$]. With these parameters, we see a $\sim$200 fold increase in CO$_2$ in the kinetics model (solid line) compared to the quench approximation model (dotted line). While this does not represent the full enhancement ($\sim$1000$\times$) we see in the PICASO best-fit models, it is meaningful increase and is therefore a strong candidate for the mechanism responsible for the CO$_2$ enhancement needed to fit our spectra.

For phosphorus, it could be that our understanding of the chemical pathways are incomplete. If the chemical timescale for PH$_3$ to be removed by oxidation or condensation (e.g. P$_4$O$_6$, P$_4$O$_{10}$, H$_3$PO$_4$, NH$_4$H$_2$PO$_4$) is shorter than is currently believed \citep{visscher_atmospheric_2006,wang_modeling_2016,morley_l_2018,visscher_mapping_2020}, the appearance of PH$_3$ absorption in model spectra would begin at higher \kzz{}'s and/or lower \teff's. Even in \wnameM, the atmospheric model is sufficiently hot that with a low \kzz{}, PH$_3$ absorption is weak and does not need to be reduced, so a small change in the phosphorus chemistry could cause the PH$_3$ features to disappear.

At a more basic level, it is possible there is no longer enough phosphorus in gaseous form to create the phosphine absorption feature. The condensate NH$_4$H$_2$PO$_4$ could act as sink for phosphorus, as discussed in \citet{visscher_atmospheric_2006} and \citet{morley_l_2018}. The condensation curve of NH$_4$H$_2$PO$_4$, assuming it condenses directly from PH$_3$, crosses the PICASO $P/T$ profiles between 1 and 0.5 bar for all our objects except for \wnameM, where it crosses at 0.1 bar. This would be consistent with out observation that the PH$_3$ mismatch gets worse for colder objects where NH$_4$H$_2$PO$_4$ is more likely to condense out. As an aside, the clear detection of PH$_3$ on Jupiter and Saturn makes the P chemistry on warmer objects that much more enigmatic. It is possible that the super-solar metallicities of these gas giants \citep{mahaffy_noble_2000,orton_vertical_2000} are responsible for these PH$_3$ detections. Observing cold metal-rich objects may prove fruitful in our search for a clear phosphine detection, and would be an good test our understand of phosphorus chemistry. Clearly the chemistry of carbon dioxide and phosphine in these ultracool atmospheres are exciting areas in need of further modeling efforts.

\section{Conclusion}
We show comparisons between three forward model grids and four NIRSpec PRISM spectra of late T and Y dwarfs, two of which are being presented for the first time in this paper. These comparisons reveal that the current generation of forward models generally overestimate the PH$_3$ absorption from 4$-$4.5\mic, and underestimate the CO$_2$ absorption in this same range. As a further test, we generate a grid of models using PICASO by taking the Elf Owl $P/T$ and chemical profiles and modify the abundances of CO$_2$ and PH$_3$, and then fit our four objects with this grid. These PICASO models show a preference for enhanced CO$_2$ abundance by at least a factor of a 100 and reduced the PH$_3$ abundance for three the coldest objects compared to the Elf Owl models. The warmest object shows minimal PH$_3$ absorption in the base best-fit Elf Owl model, and so does not need PH$_3$ reduction. We present several ways one or both of these chemical changes can be explained. These include a \kzz{} parameter that is variable with altitude, condensation of phosphorus in the form of NH$_4$H$_2$PO$_4$, and possibly an incomplete understanding of phosphorus and carbon chemical pathways. However, our favored explanation for the required CO$_2$ enhancement is that the quench approximation does not accurately predict the CO$_2$ abundance, as CO$_2$ remains in chemical equilibrium with CO after CO quenches. This results in a $\sim$200 enhancement of CO$_2$ for parameters similar to the PICASO best fits. More modeling work is needed to understand the phosphine and carbon dioxide chemistry in ultracool atmospheres.

\section{acknowledgments}
This work was inspired by conversations between some of the authors at The First Year of JWST Science Conference, and we thank the organizers for the this collaboration space. This work is based [in part] on observations made with the NASA/ESA/CSA James Webb Space Telescope. The data were obtained from the Mikulski Archive for Space Telescopes at the Space Telescope Science Institute, which is operated by the Association of Universities for Research in Astronomy, Inc., under NASA contract NAS 5-03127 for JWST. These observations are associated with program \#2302. The specific observations analyzed can be accessed via \dataset[10.17909/ntwg-k441]{https://doi.org/10.17909/ntwg-k441}. Support for program \#2302 was provided by NASA through a grant from the Space Telescope Science Institute, which is operated by the Association of Universities for Research in Astronomy, Inc., under NASA contract NAS 5-03127. This research has benefited from the Y Dwarf Compendium maintained by Michael Cushing at \url{https://sites.google.com/view/ydwarfcompendium}, and also from conversations with the Brewster retrieval community. This research has made use of the Spanish Virtual Observatory (https://svo.cab.inta-csic.es) project funded by MCIN/AEI/10.13039/501100011033/ through grant PID2020-112949GB-I00. SM and CV acknowledge support from JWST cycle 1 GO AR theory program PID-2232.

\appendix
\section{Absorption Coefficient Plots}
For completeness, we present the plots of absorption coefficients for important molecules (H$_2$O, CH$_4$, CO, NH$_3$, PH$_3$, H$_2$S, and CO$_2$) from the best-fit PICASO models for \wnameM{} and \wnameOld.
\begin{figure}
\includegraphics[width=0.5\textwidth]{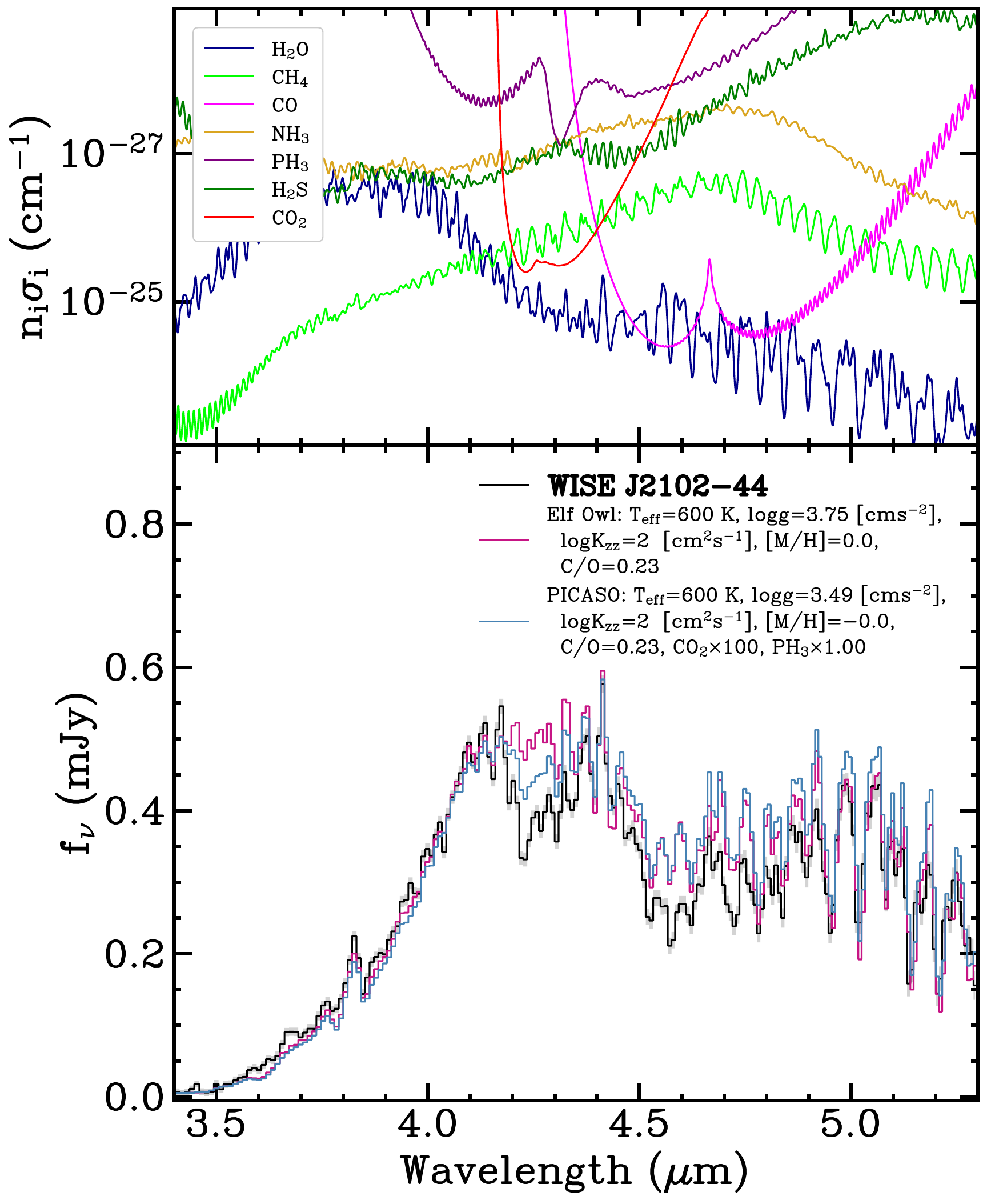}
\centering
\caption{The same plot as Figure \ref{fig:2056Picaso} but with the fits for \wnameM. The abundances are from the PICASO best-fit model at P = 1 bar, where T = 861 K, and the cross sections at T = 850 K.}\label{fig:2102Picaso}
\end{figure}

\begin{figure}
\includegraphics[width=0.5\textwidth]{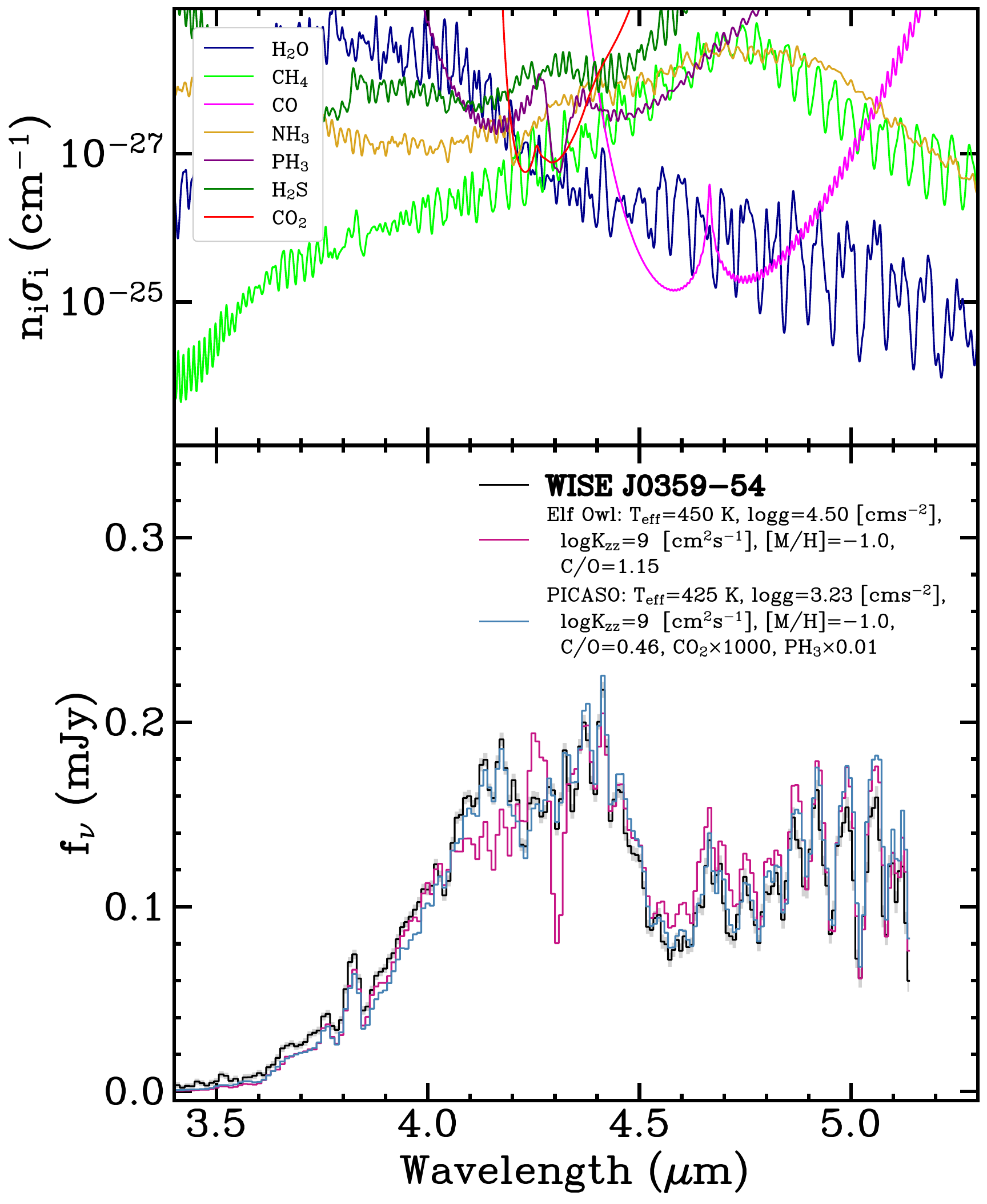}
\centering
\caption{The same plot as Figure \ref{fig:2056Picaso} but with the fits for \wnameOld. The abundances are from the PICASO best-fit model at P = 1 bar, where T = 548 K, and the cross sections at T = 550 K.}\label{fig:0359Picaso}
\end{figure}

\bibliography{sample631.bib}
\bibliographystyle{aasjournal}

\end{document}